\documentclass[conference]{IEEEtran}
\IEEEoverridecommandlockouts
\usepackage{cite}
\usepackage{amsmath,amssymb,amsfonts}
\usepackage{algorithmic}
\usepackage{graphicx}
\usepackage{textcomp}
\usepackage{hyperref}
\usepackage{xcolor}
\usepackage{multirow}
\usepackage{amsthm}

\usepackage[linesnumbered,ruled,vlined]{algorithm2e}
\def\BibTeX{{\rm B\kern-.05em{\sc i\kern-.025em b}\kern-.08em
    T\kern-.1667em\lower.7ex\hbox{E}\kern-.125emX}}

\usepackage{listings}
\definecolor{codegreen}{rgb}{0,0.6,0}
\definecolor{codegray}{rgb}{0.5,0.5,0.5}
\definecolor{codepurple}{rgb}{0.58,0,0.82}
\definecolor{backcolour}{rgb}{0.99,0.99,0.99}

\definecolor{delim}{RGB}{20,105,176}
\definecolor{numb}{RGB}{106, 109, 32}
\definecolor{string}{rgb}{0.64,0.08,0.08}

\lstdefinestyle{qfaasstyle}{
    backgroundcolor=\color{backcolour},   
    commentstyle=\color{codegreen},
    keywordstyle=\color{magenta},
    numberstyle=\tiny\color{codegray},
    stringstyle=\color{codepurple},
    basicstyle=\ttfamily\footnotesize,
    breakatwhitespace=false,         
    breaklines=true,                 
    captionpos=b,
    title={},
    keepspaces=true,                 
    numbers=none,                    
    numbersep=10pt,                  
    showspaces=false,                
    showstringspaces=false,
    showtabs=false,                  
    tabsize=1,
    frame=single,
    xleftmargin=0.1in,
    xrightmargin=0.1in
}

\lstdefinelanguage{java}{
    rulecolor=\color{black},
    postbreak=\raisebox{0ex}[0ex][0ex]{\ensuremath{\color{gray}\hookrightarrow\space}},
    upquote=true,
    morestring=[b]",
    literate=
     *{0}{{{\color{numb}0}}}{1}
      {1}{{{\color{numb}1}}}{1}
      {2}{{{\color{numb}2}}}{1}
      {3}{{{\color{numb}3}}}{1}
      {4}{{{\color{numb}4}}}{1}
      {5}{{{\color{numb}5}}}{1}
      {6}{{{\color{numb}6}}}{1}
      {7}{{{\color{numb}7}}}{1}
      {8}{{{\color{numb}8}}}{1}
      {9}{{{\color{numb}9}}}{1}
      {\{}{{{\color{delim}{\{}}}}{1}
      {\}}{{{\color{delim}{\}}}}}{1}
      {[}{{{\color{delim}{[}}}}{1}
      {]}{{{\color{delim}{]}}}}{1},
}

\newcommand*{\code}{\lstinline[keywordstyle=\color{blue}, basicstyle=\ttfamily\small\color{black}]}

\begin{document}

\title{QSimPy: A Learning-centric Simulation Framework for Quantum Cloud Resource Management
}

\author{
    \IEEEauthorblockN{Hoa T. Nguyen\textsuperscript{1}, Muhammad Usman\textsuperscript{2,3}, and Rajkumar Buyya\textsuperscript{1}}
    \IEEEauthorblockA{\textsuperscript{1}\textit{Cloud Computing and Distributed Systems (CLOUDS) Laboratory}, \\ \textit{School of Computing and Information Systems, The University of Melbourne, Parkville, 3052, Victoria, Australia}}
    \IEEEauthorblockA{\textsuperscript{2}\textit{School of Physics, The University of Melbourne, Parkville, 3052, Victoria, Australia}}
    \IEEEauthorblockA{\textsuperscript{3}\textit{Data61, CSIRO, Clayton, 3168, Victoria, Australia}} 
    thanhhoan@student.unimelb.edu.au, \{muhammad.usman, rbuyya\}@unimelb.edu.au
    
}

\maketitle

\begin{abstract}
Quantum cloud computing is an emerging computing paradigm that allows seamless access to quantum hardware as cloud-based services. However, effective use of quantum resources is challenging and necessitates robust simulation frameworks for effective resource management design and evaluation. To address this need, we proposed QSimPy, a novel discrete-event simulation framework designed with the main focus to facilitate learning-centric approaches for quantum resource management problems in cloud environments. Underpinned by extensibility, compatibility, and reusability principles, QSimPy provides a lightweight simulation environment based on SimPy, a well-known Python-based simulation engine for modeling dynamics of quantum cloud resources and task operations. We integrate the Gymnasium environment into our framework to support the creation of simulated environments for developing and evaluating reinforcement learning-based techniques for optimizing quantum cloud resource management. The QSimPy framework encapsulates the operational intricacies of quantum cloud environments, supporting research in dynamic task allocation and optimization through DRL approaches. We also demonstrate the use of QSimPy in developing reinforcement learning policies for quantum task placement problems, demonstrating its potential as a useful framework for future quantum cloud research.

\end{abstract}


\section{Introduction}
\begin{figure}[htbp]
\centering
\includegraphics[width=3.3in]{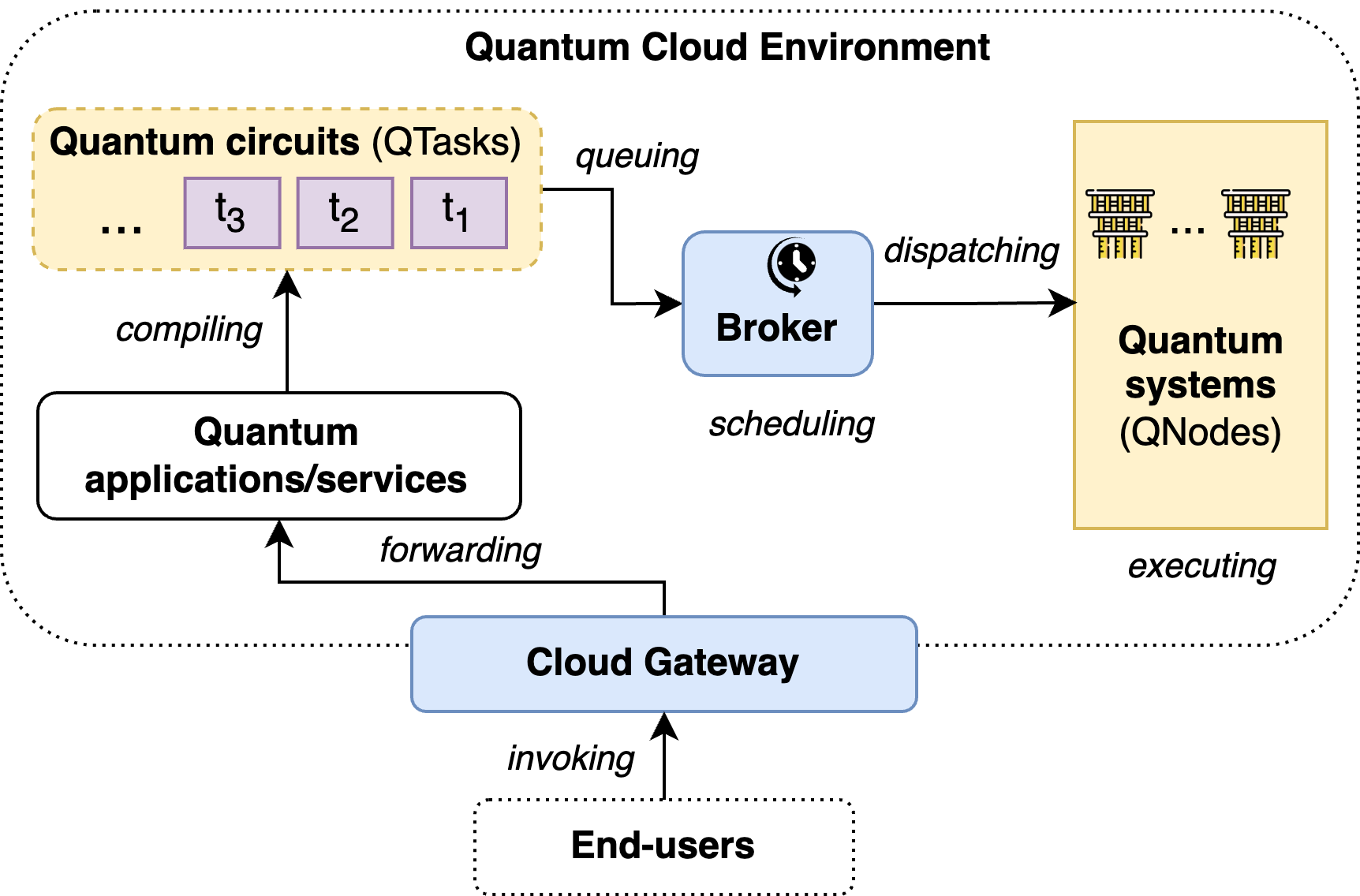}
\caption{A high-level view of quantum cloud computing environments and a sample procedure from user invocation to quantum task execution.}
\label{fig_overview}
\end{figure}

Quantum computing, characterized by its capacity to tackle impractically complex computation tasks for classical systems, offers revolutionary potential across multiple domains, such as drug discovery \cite{zinner2021quantum}, machine learning \cite{zhang2020recent}, and optimization \cite{lewis2017quadratic}. However, this emerging paradigm faces significant practical barriers, notably the operational challenges and substantial costs associated with operating quantum hardware \cite{de_leon2021materials}. The advent of quantum cloud computing \cite{nguyen2024quantum,n2023quantum} offers a pathway to overcome these barriers, providing access to quantum computing capabilities through cloud services. Figure \ref{fig_overview} illustrates the overview of the contemporary quantum cloud computing environment and an example of how end-users interact with cloud-based quantum applications or services. 
Currently, classical systems are utilized to host these quantum applications \cite{ravi2021adaptive}, which define the quantum execution logic using a quantum software development kit (SDK), such as Qiskit \cite{contributors2023qiskit}. Typically, end-users can manually build and compile quantum circuits within a cloud-based integrated development environment, such as Jupyter Notebook, and submit them for execution in a quantum system. Another comprehensive approach involves deploying quantum applications or functions as cloud-based services, offering API endpoints for external invocation \cite{nguyen2024qfaas}. Through this service model, the user can invoke a quantum service via a cloud gateway; then, a corresponding quantum circuit is generated and compiled. These compiled circuits are then enqueued within a broker, which is responsible for the task scheduling and the subsequent dispatch to the selected quantum system for execution.

% Resource management challenges
Despite the benefits, integrating quantum computing into cloud environments introduces complex resource management challenges. These challenges arise from the unique properties of quantum computations, such as the different execution models compared to classical computation tasks \cite{nguyen2024qfaas} and the current limitations of Noisy intermediate-scale quantum (NISQ) hardware \cite{preskill2018quantum}. Additionally, the heterogeneous nature of quantum cloud resources \cite{ravi2021adaptive}, varying in architectural design and capability, complicates the effective allocation and scheduling of quantum tasks.

% Limitations of the literature - gaps
Current literature primarily focuses on quantum algorithms and physical quantum operation simulations \cite{serrano2022quantum}, with less emphasis on the holistic simulation of quantum cloud environments that include both hardware and service management aspects. Most existing quantum simulators, such as Qiskit Aer \cite{contributors2023qiskit}, or cloud simulators, such as CloudSim \cite{calheiros2011cloudsim}, do not support the modeling of quantum cloud computing environments. In this domain, our prior work, iQuantum \cite{nguyen2024iquantum}, stands as a pioneering Java-based framework utilizing CloudSim for modeling quantum cloud environments. iQuantum has laid a critical foundation; however, its Java-centric nature presents challenges when interfacing with the predominantly Python-based quantum software and machine learning libraries, despite potential Java-Python interoperability solutions such as Py4J \cite{j2024py4j}. Given the predominance of Python in the quantum software landscape \cite{serrano2022quantum}, as seen in SDKs like Qiskit and Cirq and machine learning libraries like TensorFlow, PyTorch, and Ray, the demand for a Python-native simulation tool is evident. Furthermore, the specific demands of machine learning integration for dynamic resource management in quantum cloud computing pose a steep learning curve for those who are unfamiliar with CloudSim-base methodologies in the cloud computing domain.

\begin{table*}[htbp]
\caption{Representative works related to our study}
\centering
\label{tab:related-work}
\begin{tabular}{|l|l|l|l|l|l|}
\hline
\textbf{Work} & \textbf{\begin{tabular}[c]{@{}l@{}}Simulation \\ Approach\end{tabular}} & \textbf{Focus} & \textbf{\begin{tabular}[c]{@{}l@{}}Programming\\ Language\end{tabular}} & \textbf{\begin{tabular}[c]{@{}l@{}}ML \\ Integration\end{tabular}} & \textbf{Dataset Support} \\ \hline
qgym \cite{linde2023qgym} & - & \begin{tabular}[c]{@{}l@{}}Quantum compiler\\ benchmarking\end{tabular} & Python & Gymnasium & - \\ \hline
DRAS-CQSim \cite{fan2021dras} & Discrete-event & \begin{tabular}[c]{@{}l@{}}RL-based HPC \\ cluster scheduling\end{tabular} & Python & \begin{tabular}[c]{@{}l@{}}Tensorflow,\\ Keras\end{tabular} & - \\ \hline
Jawaddi et al. \cite{agosjawaddi2024integrating} & Discrete-event & \begin{tabular}[c]{@{}l@{}}RL-based Classical \\ Cloud Environments\end{tabular} & Java & \begin{tabular}[c]{@{}l@{}}Open AI Gym,\\ Keras\end{tabular} & - \\ \hline
iQuantum \cite{nguyen2024iquantum} & Discrete-event & \begin{tabular}[c]{@{}l@{}}Quantum Cloud-Edge\\ Environments\end{tabular} & Java & - & CSV \\ \hline
\textit{\textbf{QSimPy (This work)}} & Discrete-event & \begin{tabular}[c]{@{}l@{}}RL-based Quantum \\ Resource Management\end{tabular} & Python & \begin{tabular}[c]{@{}l@{}}Gymnasium, \\ Ray\end{tabular} & QASM, CSV \\ \hline
\end{tabular}
\end{table*}

To address these considerations and close the gap, we propose QSimPy, a Python-based simulation framework specifically tailored for creating learning-centric environments for modeling and evaluating resource management strategies in quantum cloud computing. QSimPy is not only the Python-based counterpart of iQuantum but also represents a starting point for the paradigm shift toward a more expansive, integrated simulated environment compatible with well-known frameworks in both the quantum computing and machine learning ecosystems. Our framework aims to facilitate the development and evaluation of resource management policies using reinforcement learning (RL) techniques. It models not only quantum computation but also resource orchestration in a cloud-based quantum computing environment, enabling the simulation of various scenarios and policies to optimize resource utilization effectively. By offering a learning-centric environment, QSimPy allows researchers and practitioners to design and evaluate RL-based strategies for resource management, addressing both the performance variability of quantum devices and the fluctuating demands of quantum computing tasks. This simulation tool is developed to advance quantum cloud resource management by offering a platform for prototyping, modeling and conducting experimentation, ultimately enabling the development of more efficient and robust resource management strategies for quantum cloud computing.

\subsection{Main Contributions}
This paper proposes QSimPy, a Python-based simulation framework that leverages the SimPy \cite{y2024simpy} discrete-event simulation approach to focus on reinforcement learning techniques for resource management in quantum cloud environments. The main contributions and novel aspects of our work are summarized as follows:
\begin{itemize}
    \item We present the system model specifically designed for the quantum cloud resource management problem, tailored for the reinforcement learning approaches. This model serves as a foundational hypothetical basis for simulating and analyzing the dynamics of quantum task execution in cloud environments.
    \item We propose QSimPy, a Python-based discrete-event simulation framework, leveraging the extensive ecosystem of Python libraries tailored for machine learning and quantum computing. QSimPy offers a flexible and powerful environment for developing complex resource management algorithms with machine learning techniques. Its design focuses on ease of integration with external libraries, making it a potential starting point for collaborations between researchers and practitioners in quantum and cloud computing to enhance quantum cloud resource management.
    \item We demonstrate QSimPy’s capabilities through its application to a quantum task placement problem. This proof of concept implementation showcases how QSimPy integrates seamlessly with industry-grade RL environments and algorithm libraries, such as Gymnasium and Ray RLlib. Our results confirm the framework's functionality and potential as a benchmark tool for future quantum cloud computing resource management research.
    
\end{itemize}

These contributions highlight QSimPy's innovative aspects and its potential to significantly impact the field of quantum cloud computing. By integrating state-of-the-art RL methodologies with a flexible simulation environment, QSimPy offers a novel tool that enhances the study and practical management of cloud-based quantum computing resources.

\subsection{Paper Organization}
The rest of the paper is organized as follows: Section II reviews the relevant literature, identifying the research gaps that our study aims to bridge. Section III outlines the system model and problem formulation for resource management within a quantum cloud computing environment tailored to reinforcement learning approaches. Section IV details the design, primary components, and implementation of the QSimPy framework. Section V demonstrates an exemplary application of our framework, implementing RL-based task placement policies in a heterogeneous quantum cloud environment. Section VI concludes the paper by summarizing our key findings and outlining future research directions.

\section{Related Work}
This section presents a brief review of related work, illustrating the unique gap our research addresses in the domain of quantum cloud computing simulation. We summarise several representative works and precedent studies in Table \ref{tab:related-work}.

In the high-performance computing (HPC) cloud computing domains, DRAS-QCsim \cite{fan2021dras} employs reinforcement learning for efficient cluster scheduling. However, it does not specifically address quantum cloud environments. Other endeavors, such as the work of Jawaddi et al. \cite{agosjawaddi2024integrating}, explore resource management within classical cloud environments using reinforcement learning yet do not cross the bridge into supporting the quantum computing domain. 

In the domain of quantum computing simulations, a variety of frameworks exist, each with its distinct focus and methodology. For example, qgym \cite{linde2023qgym} focuses on quantum compiler benchmarking without delving into resource management of the quantum cloud environment. Considering the modeling and simulation of quantum cloud environments, our prior work, iQuantum \cite{nguyen2024iquantum}, can be considered as one of the first-of-its-kind frameworks. iQuantum is a Java-based modeling and simulation framework that initiates the quantum cloud simulation, based on the core simulation engine of CloudSim \cite{calheiros2011cloudsim} but with a specific focus on quantum cloud-edge computing environment modeling and simulation without considering the integration of machine learning or reinforcement learning techniques for optimization. 

Our proposed QSimPy aims to fill the noticeable gap in the current landscape of simulation software. It stands out by adopting a discrete-event approach tailored to the intricate problem of quantum cloud resource management through reinforcement learning approaches. With QSimPy, we have developed a Python-based framework that is not only compatible with the massive ecosystem of software development kits, libraries, and other frameworks in quantum computing and machine learning but also simplifies the usage, customization, and further extension for more complex quantum cloud environment modeling and simulation. Unlike its predecessors, such as CloudSim \cite{calheiros2011cloudsim} and iQuantum \cite{nguyen2024iquantum}, QSimPy is designed to be compatible with industry-graded machine learning environments and libraries, such as Gymnasium and Ray RLlib, facilitating the implementation and validation of advanced resource management strategies. Our framework is also built to support a variety of datasets, including OpenQASM \cite{cross2017open} quantum circuits and CSV-based quantum circuit features, ensuring its versatility and applicability across various synthetic quantum computing tasks dataset generation. Our framework not only addresses the limitations of existing simulators but also pioneers a pathway for future studies and developments in quantum cloud resource management algorithms. By bridging the gap between classical and quantum resource orchestration, QSimPy serves as a potential starting point for event-based simulation and optimization of quantum tasks within cloud environments, a step towards further investigation to harness the full potential of quantum cloud computing. 

\begin{figure*}[htbp]
\centering
\includegraphics[width=6.4in]{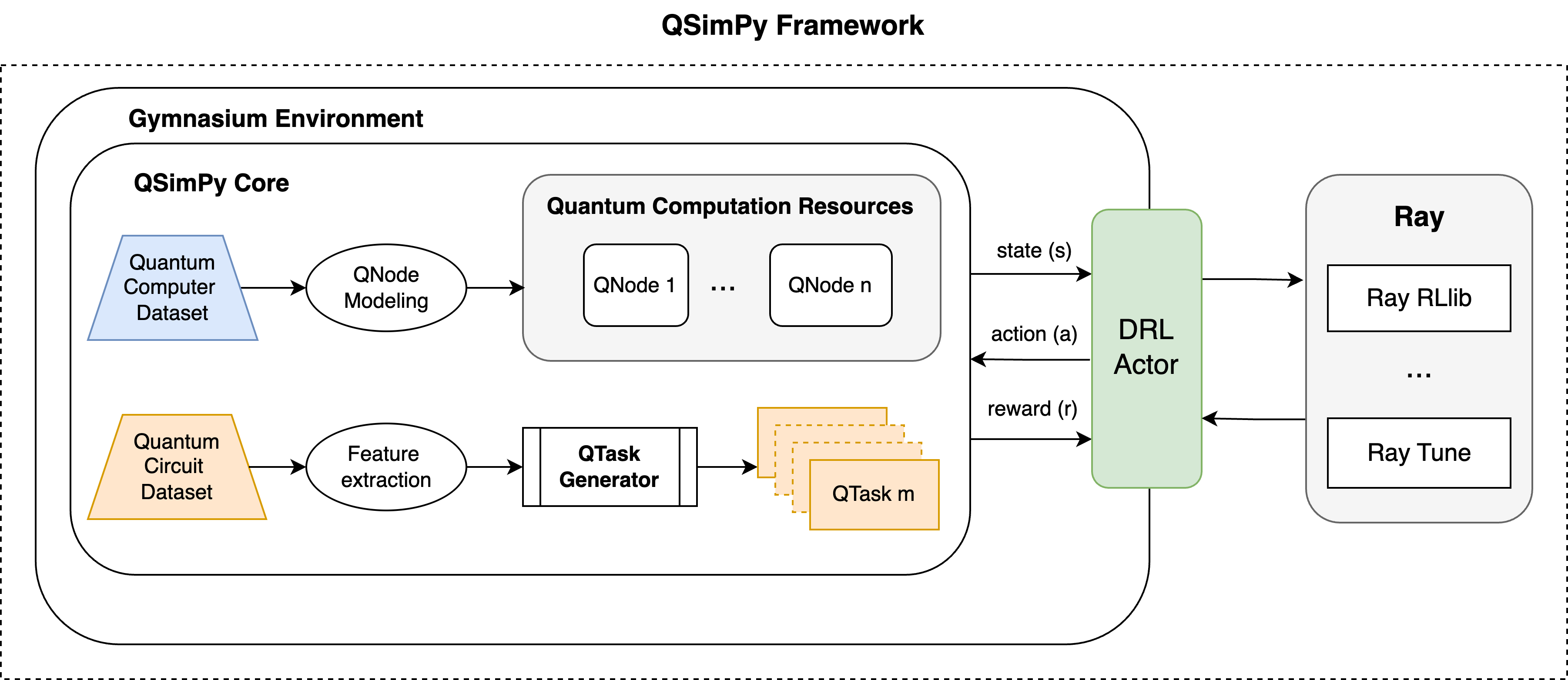}
\caption{An overview of the QSimPy Framework. }
\label{fig_qsimpy}
\end{figure*}

\section{RL-based Problem Formulation and Model}
\subsection{Problem Formulation}
The quantum cloud resource management (QCRM) problem can be defined as the process of determining the optimal policy to place incoming quantum tasks in the best-suited quantum computation resource so that the optimization objective can be achieved successfully. This objective can vary from different criteria, including 1) time-awareness (such as minimizing the total task completion time), 2) precision-awareness (such as minimizing the error rates in the task execution results), and 3) utilization-awareness (such as maximizing the overall resource utilization). A complex problem can also take multiple objectives into account at the same time.

\subsection{Reinforcement Learning-based QSTAR Model}

Let QCRM problem $\mathcal{M}$ in the reinforcement learning setting be a set of components denoted as a QSTAR model as follows:
\begin{equation}
    \mathcal{M} = \{\mathbb{Q}, \mathbb{S}, \mathbb{T}, \mathbb{A},  \mathbb{R} \}
\end{equation}
where:
\begin{enumerate}
    \item \textbf{Quantum Resources:} $\mathbb{Q} = \{Q_1, Q_2, ..., Q_n\}$ denoted the list of $n$ quantum cloud computation resources, or quantum nodes (QNodes). Each QNode $Q$ can also be modeled as $Q = \{q_1, q_2, ..., q_k\}$ where $q_1,...,q_k$ are $k$ different computation capacity metrics such as the number of qubits, quantum volume \cite{cross2019validating}, circuit layer operation per seconds (CLOPS) \cite{wack2021quality}, native quantum gates, qubit connectivity, and error rates. 
    \item \textbf{Quantum Tasks:} $\mathbb{T} = \{T_1, T_2, ..., T_m\}$ denotes the list of $m$ quantum tasks (QTask) that need to be executed on one of the available quantum nodes. Each QTask $T$ can be presented as $T = \{t_1, t_2, ..., t_l\}$ where $t_1,...,t_l$ are $l$ different attributes of QTask such as the number of qubits, circuit layers, required quantum gates, qubit connectivity, and other quality of service (QoS) metrics.
    \item \textbf{State space:} $\mathbb{S}$ is all possible state of the agent's observation from the quantum cloud environment, which includes 1) information of all available quantum resources (QNodes) and 2) information of current QTasks to be scheduled.
    The feature vector of $n$ quantum nodes $Q_i \in \mathbb{Q}$, each quantum node has $k$ features, at time step $\tau \in \mathcal{T}$ can be presented as:
    \begin{equation}
        \mathcal{F}^{\mathbb{Q}}_\tau = \{f^{q_z}_{Q_i} | \forall Q_i \in \mathbb{Q}, 1 \le i \le n, 1 \le z \le k \}
    \end{equation}
    where $i$ is the index of the quantum node, and $z$ is the index of the quantum node's metric.
    
    The feature vector of current quantum task (QTask) $T_j \in \mathbb{T}$ need to be scheduled, which has  $l$ features at time step $\tau \in \mathcal{T}$ can be presented as:
    \begin{equation}
    \mathcal{F}^{T_j}_\tau = \{f^{t_p}_{T_j} |  T_j \in \mathbb{T}, 1 \le j \le m, 1 \le p \le l \}
    \end{equation}
    where $j$ is the index of the current quantum task, and $p$ is the index of the current quantum task's metric.

    Therefore, the state space $\mathbb{S}$ of the overall system can be represented as follows:
    \begin{equation}
        \mathbb{S} = \{s_\tau | s_\tau = (\mathcal{F}^{\mathbb{Q}}_\tau, \mathcal{F}^{T_j}_\tau), \forall \tau \in \mathcal{T}, T_j \in \mathbb{T} \}
    \end{equation}
    \item \textbf{Action space:} $\mathbb{A}$ is all possible actions that can be taken without the quantum cloud environment. An action can be defined as a mapping of a QTask to an available QNode. The action $a_\tau$ at time step $\tau$ is an placement of QTask $T_j$ to quantum node $Q_i$ can be defined as
    \begin{equation}
        a_\tau = \{Q_i, T_j\}, Q_i \in \mathbb{Q}, T_j \in \mathbb{T}, 
    \end{equation}
    Thus, the Action space is equivalent to the set of all available quantum nodes at the data center:
    \begin{equation}
        \mathbb{A} = \mathbb{Q}
    \end{equation}

    \item \textbf{Reward function} $\mathbb{R}$ determine the reward or penalty that assigned to each action $a_\tau$. A sample reward function can be defined as follows:
    \begin{equation}
    \label{reward}
        r_\tau = 
    \begin{cases} 
    \delta^+ & \text{if } \text{success} = 1 \\
    \delta^- & \text{if } \text{success} = 0 
    \end{cases}
    \end{equation}
    where $\delta^+$ is the reward (typically a positive number) applied when the QTask can be successfully executed, and $\delta^-$  and is the penalty (typically a negative number) applied if the QTask cannot be executed if the action violates the predefined constraints of the problem. This is to guide the agent to avoid taking similar action in the future.

\end{enumerate}

\section{The QSimPy Framework}
\subsection{Design Overview}
The design of our QSimPy simulation framework is driven by the following principles:
\begin{itemize}
    \item \textbf{Extensibility}: QSimPy is designed to seamlessly accommodate future enhancements and extensions. By adopting a modular design approach, each framework component is designed to be easily extendable, allowing researchers and practitioners to incorporate new features or functionalities without compromising the integrity of the existing framework. This extensibility ensures that QSimPy remains adaptable to evolving research needs and technological advancements in the field of quantum hardware and quantum cloud computing.
    \item \textbf{Compatibility}: This is a cornerstone of the QSimPy framework, ensuring interoperability with existing tools, libraries, and standards commonly utilized in the quantum computing and machine learning communities. QSimPy is designed to integrate easily with industry-standard machine learning environments and libraries, such as Gymnasium and Ray RLlib, facilitating the adoption and validation of advanced resource management strategies. Additionally, QSimPy supports a variety of datasets, including OpenQASM and CSV, enhancing its compatibility with diverse quantum computing tasks and research methodologies.
    \item \textbf{Reusability}: QSimPy emphasizes reusability by providing a flexible and versatile framework for quantum cloud simulation and resource management. The framework's modular design encourages the reuse of components across different simulation scenarios and research projects, promoting efficiency and reducing development overhead.
    
\end{itemize}

\subsection{Architecture Design and Functionality}
\subsubsection{Core Simulation Engine}
The core simulation engine of QSimPy is built on top of the robust foundation provided by SimPy \cite{y2024simpy}, a well-established and versatile discrete-event simulation Python library. A discrete-event simulation is an approach that models the operation of a system as a sequence of events in time. Each event occurs at an instant in time and marks a change of state in the system \cite{goldsman2015discrete}. By leveraging SimPy, QSimPy inherits a proven framework for representing complex, stochastic systems where interdependent events drive the evolution of system states. This approach is particularly advantageous for modeling quantum cloud environments, as it allows for the proper representation of asynchronous and concurrent operations within quantum computation resources. 

\subsubsection{Dataset Generation and Modeling} 
\begin{figure}[htbp]
\centering
\includegraphics[width=2.7in]{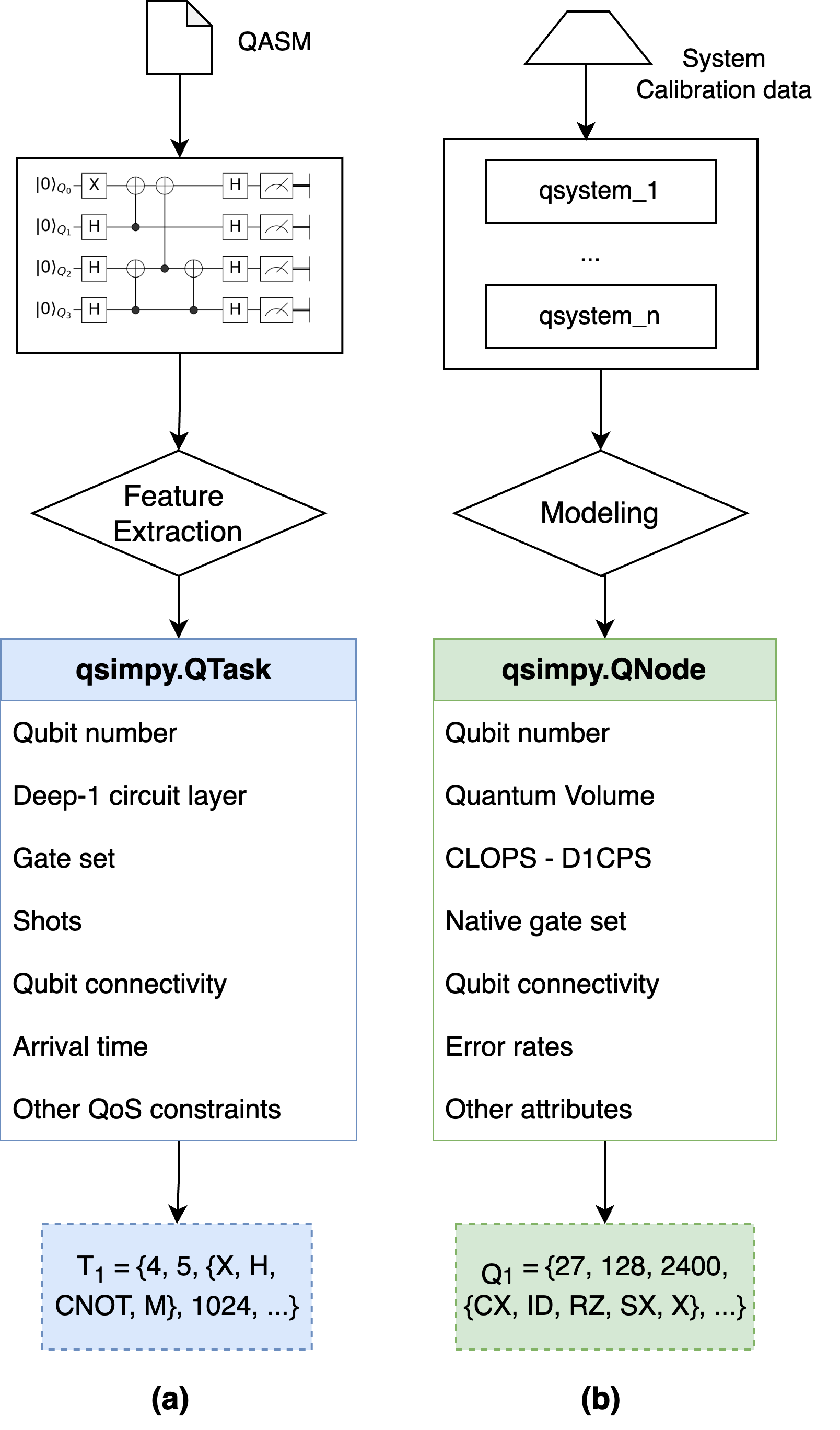}
\caption{Feature extraction and modeling for (a) quantum computation tasks (QTasks) and (b) quantum computation nodes (QNodes) in QSimPy}
\label{fig_dataset}
\end{figure}

As the quantum cloud workload dataset is not yet available, generating a reliable synthetic dataset is essential for designing and evaluating quantum cloud resource management policies. Figure \ref{fig_dataset} illustrates the modeling and dataset generation process. We develop a QTask dataset generation module for QSimPy by extracting circuit features from the MQTBench \cite{quetschlich2023mqt} quantum circuit dataset and mapping with other QTask features, such as arrival time and shots (task execution repetition) based on the defined distribution. We also use quantum system calibration data from quantum cloud providers (such as IBM Quantum) for modeling quantum node instances.

\subsubsection{Gymnasium-based Environment} We have devised an environment wrapper utilizing Gymnasium \cite{towers_gymnasium_2023}, a leading API standard in the reinforcement learning domain. This integration enables the simulation framework to establish an interface for reinforcement learning algorithms, allowing them to interact with quantum cloud computing environments. Gymnasium standardized API facilitates the consistent development and benchmarking of RL agents within our simulation platform.

\subsubsection{RL Framework Connector} 
The Gymnasium-compliant QSimPy environment paves the way for interoperability with a suite of state-of-the-art reinforcement learning frameworks. Prominent among these are Ray RLlib \cite{liang2018rl}, Stable Baselines 3 \cite{raffin2021stable}, and Tianshou \cite{weng2022tianshou}, which serve as popular tools for implementing and training RL agents. We incorporate the Ray ecosystem to further strengthen QSimPy's extensibility and scalability, specifically leveraging the Ray RLlib for RL algorithm implementation and Ray Tune \cite{liaw2018tune} for systematic hyperparameter optimization. This approach ensures that our framework facilitates cutting-edge RL-based quantum resource management strategies.

\subsection{Simulation Workflow}
The simulation workflow of QSimPy, as outlined in Algorithm \ref{alg:qsimpy_simulation_logic}, orchestrates the interaction between quantum task execution and resource management through a discrete-event simulation process.

\begin{algorithm}[htbp]
\SetAlgoLined
\caption{QSimPy Simulation Logic}\label{alg:qsimpy_simulation_logic}
Initialize QSimPy Gymnasium environment\;
Instantiate QSimPy computation resources (QNodes)\;
Generate quantum computing tasks (QTasks)\;
Instantiate QBroker for managing resources and tasks\;
Connect RL Actor to the environment for decision-making \;
\For{each QTask in the system}{
    Sends QTask to QBroker\;
    QBroker collects observations from the environment\;
    QBroker sends the observation to the RL Actor\;
    RL Actor performs training and returns an action\;
    QBroker dispatched QTask to a QNode based on the action\;
    \eIf{QNode is available}{
        Generate the execution event\;
        Update the state of QNodes\;
    }{
        Queue QTask in QNode for later execution\;
    }
    QBroker collects observations from the environment\;
    QBroker sends the observation to the RL Actor\;
    RL Actor calculates reward and improve the policy\;
}
\end{algorithm}

The simulation initiates by establishing the Gymnasium environment, which acts as a standardized interface for reinforcement learning agents. Then, computational resources, QNodes, are instantiated to simulate the quantum computing resources. Besides, quantum computing tasks (QTasks), are generated and fed into the system, emulating the workload of quantum computations. A QBroker component is responsible for the allocation and management of these task placements. It serves as an intermediary between the QTasks and the QNodes, observing the state of the environment and delegating tasks accordingly. The RL Actor, a reinforcement learning agent, is connected to the simulation for dynamic decision-making. It receives the environment's state from the QBroker and computes an action that dictates the subsequent allocation of the QTasks. These actions are based on the objective of optimizing resource utilization or other resource management objectives within the quantum cloud environment.

After the action determination from the RL Actor, the QBroker dispatches QTasks to QNodes, delegation on their availability. If a QNode is ready for task execution, the corresponding event is generated, and the QNode state is updated. Otherwise, the QTask is queued, awaiting future execution. The simulation proceeds iteratively, with the QBroker continuously collecting observations and interacting with the RL Actor, which refines its policy based on the received state information and the rewards calculated from the actions taken. This iterative process underpins the RL agent's ability to learn and adapt its task placement strategy over time, striving for an optimized management of quantum computational resources.

The use of reinforcement learning within our simulation framework introduces a layer of learning-centric adaptability, enabling the system to improve its resource allocation decisions through continuous interaction with the quantum cloud environment. This approach encapsulates the dynamism and unpredictability inherent in quantum cloud resource management, providing a simulated environment for researchers to explore and develop scheduling and orchestrating algorithms.

\subsection{Implementation}
QSimPy is implemented as a Python package to facilitate the environmental setup of reinforcement learning-based quantum cloud resource management problems. The core simulation engine of QSimPy is developed based on SimPy, a well-known Python library for discrete-event simulation. All environmental implementations use the Gymnasium package and comply with the most popular Reinforcement learning frameworks, such as stable-baselines3, Tianshou, and Ray RLlib. We mainly test and suggest using Ray RLlib for the reinforcement learning algorithm implementation. The implementation of QSimPy can be accessed on GitHub at \url{https://github.com/Cloudslab/qsimpy}.

\section{Example of using QSimPy}
This section discusses an example to demonstrate how to set and use the QSimPy environment step-by-step, followed by an exemplary result to show the potential use of the reinforcement learning approach for quantum task placement over heuristic approaches.

\subsection{Example scenario: DRL-based Quantum Task Scheduling}
We consider a quantum cloud data center comprising of five QNodes, with qubit numbers, varied from 7 to 127, and the qubit topologies follow five IBM Quantum systems \cite{qcloud-ibm}, including ibm\_washington, ibm\_kolkata, ibm\_hanoi, ibm\_perth, and ibm\_lagos. The detailed specifications of all QNodes are depicted in Table \ref{tab:qnode}. 
The model of QNode in our study adopts four benchmarking metrics for a quantum system as proposed by Wack et al. \cite{wack2021quality}. These metrics include the number of qubits, the quantum volume (QV), the rate of circuit layer operations per second (CLOPS), and the depth-1 circuit operations per second (D1CPS). For the purposes of our simulations, we adopt the D1CPS values for each QNode from the empirical evaluations presented in \cite{wack2021quality}. These values are used to estimate the execution time for each incoming QTask, with the calculation being contingent on the number of depth-1 circuit layers that the QTask encompasses.

\begin{table}[htbp]
\centering
\caption{QNode specifications based on IBM Quantum systems}
\label{tab:qnode}
\begin{tabular}{|l|l|l|l|l|}
\hline
\multicolumn{1}{|c|}{\textbf{QNode Model}} & \multicolumn{1}{c|}{\textbf{Qubits}} & \multicolumn{1}{c|}{\textbf{QV}} & \multicolumn{1}{c|}{\textbf{CLOPS}} & \multicolumn{1}{c|}{\textbf{D1CPS}} \\ \hline
washington & 127 & 64 & 850 & 16967.5 \\ \hline
kolkata & 27 & 128 & 2000 & 39900 \\ \hline
hanoi & 27 & 64 & 2300 & 45935 \\ \hline
perth & 7 & 32 & 2900 & 57905 \\ \hline
lagos & 7 & 32 & 2700 & 53865 \\ \hline
\end{tabular}
\end{table}

For workload, we generated the synthetic dataset of 49000 QTasks based on quantum circuits extracted from 12 quantum algorithms/applications of the MQTBench dataset \cite{quetschlich2023mqt} with the qubit number varied from 2 qubits to 27 qubits as follows:
\begin{enumerate}
    \item Amplitude Estimation
    \item Deutsch-Jozsa
    \item Greenberger-Horne-Zeilinger (GHZ) state
    \item Quantum Fourier Transformation (QFT)
    \item Entangle QFT
    \item Quantum Neural Network
    \item Quantum Phase Estimation exact
    \item Quantum Phase Estimation inexact
    \item Random Circuit
    \item Real Amplitudes ansatz with Random Parameters
    \item Efficient SU2 ansatz with Random Parameters
    \item Two Local ansatz with random parameters
\end{enumerate}
All QTasks are divided into 1900 subsets. Each subset contains random circuits of all the applications above to simulate the stochastic nature of quantum cloud computing environments. To simplify the demonstration purpose, we assume that an average of 25 QTasks arrive every minute, with the arrival time following the uniform distribution.

\textbf{Problem statement:} \textit{Given all quantum computation resources (QNodes) and continuous incoming QTasks as aforementioned, design a reinforcement learning policy to optimize the quantum task scheduling by minimizing the total completion time of incoming tasks, in which task completion time is the sum of waiting time and execution time.}

\subsection{Setting up quantum cloud environment using QSimPy}
The structure of the source code for creating the Gymnasium environment using QSimPy is as depicted in Code \ref{lst:gym-env}.

\begin{lstlisting}[language=python, caption=Sample structure for the Gymnasium-based QSimPy environment, label={lst:gym-env}]
import gymnasium as gym
from qsimpy import QTask, QNode, Broker, Dataset
...
class QSimPyEnv(gym.Env):
  def __init__(self, config, dataset):
    ...[environment initialization]

  def reset(self, *,seed, option):
    ...[reset operation]

  def step(self, action):
    ...[step definition based on action]
    ...[reward calculation]
    ...[terminate/done condition]
    
  def close(self):
    ...[other close operation]
\end{lstlisting}

To model the quantum computation resources (QNodes), users can use our predefined QNode model based on IBM Quantum system information as Code \ref{lst:qnodes}. User can also define their own customized QNode configuration using the QNode object in the QSimPy package.

\begin{lstlisting}[language=python, caption=Sample code snippet for model quantum nodes and broker, label={lst:qnodes}]
def setup_quantum_resources(self):
  qnode_ids = range(self.n_qnodes)
  qnode_names = [ "washington", "kolkata", "hanoi", "perth", "lagos",]
  self.qnodes = [IBMQNode.create_ibmq_node(self.qsp_env, qid, qname) for qid, qname in zip(qnode_ids, qnode_names)]
  self.broker = Broker(self.qsp_env, self.qnodes)
  ... [truncated]
\end{lstlisting}

Then, we can generate incoming QTasks based on the workload dataset (as depicted in Code \ref{lst:qtasks}) and define the QTask submission function (as depicted in Code \ref{lst:submit-qtasks}) to connect with QSimPy engine to simulate all the execution events.

\begin{lstlisting}[language=python, caption=Sample code snippet for generating QTasks, label={lst:qtasks}]
def generate_qtasks(self):
  qtasks = self.qdataset.get_subset(self.round)
  self.qtasks = [QTask(id, arrival_time, qtask_data) for id, arrival_time, qtask_data in zip(qtask_ids, qtask_arrival, qtask_values)]
  ... [truncated]
\end{lstlisting}

\begin{lstlisting}[language=python, caption=Sample code snippet for submitting a QTask to a QNode, label={lst:submit-qtasks}]
def submit_task_to_qnode(self, qtask, qnode_id):
  qtask_execution = self.broker.submit_qtask_to_qnode(qtask, self.qnodes[qnode_id])
  self.qsp_env.process(qtask_execution)
  ...[truncated]
\end{lstlisting}

\begin{figure}[htbp]
\centering
\includegraphics[width=3.1in]{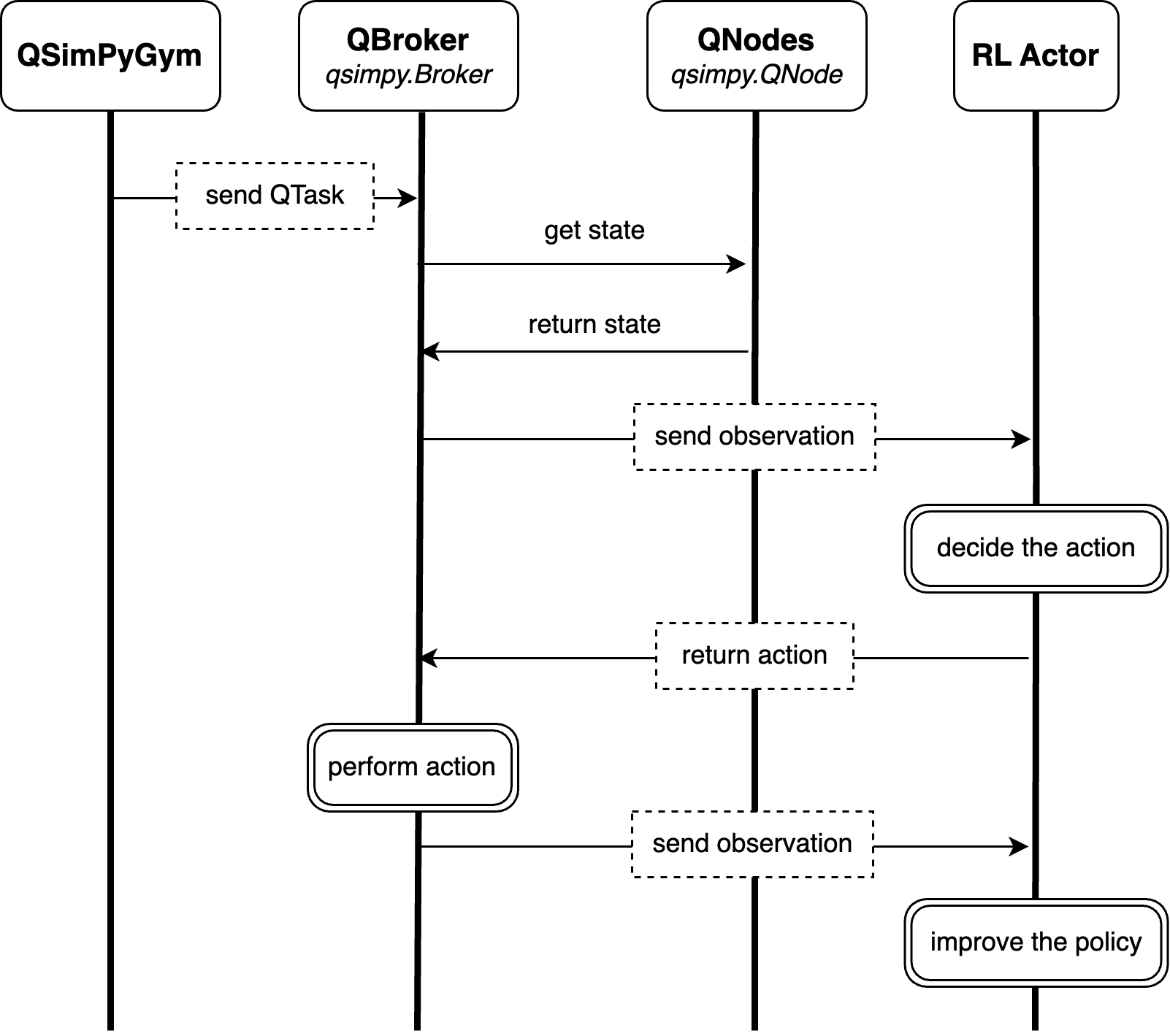}
\caption{The sequence of main steps in the action decision-making process for each incoming QTask within the QSimPy simulation environment.}
\label{seq_diagram}
\end{figure}

To achieve the task completion time minimization objective, we can define a simple reward function as follows:
\begin{equation}
\label{reward}
    r_\tau = 
\begin{cases} 
\dfrac{1}{time(T_i)} & \text{if } \text{success} = 1 \\
-10 & \text{if } \text{success} = 0 
\end{cases}
\end{equation}
where $time(T_i)$ is the total completion time of QTask $T_i$ and $-10$ is penalty value. If QTask execution is successful, we apply the reward as the inverse value of its completion time, as the reinforcement learning algorithm will try to maximize the reward. Otherwise, if the execution fails, we apply the penalty with a large negative value (for example, -10) to urge the agent to avoid taking similar action later on. This reward function is defined in the \code{step} function of the environment based on the result of the \code{submit_task_to_qnode} function. Besides, users can define a customized environment wrapper to normalize and rescale the observation and reward if needed, leveraging the features of Gymnasium. Figure \ref{seq_diagram} illustrates the main steps involved in the simulation process of submitting each incoming QTask to QSimPy during the RL training.

First, a QTask object is generated with attributes from the quantum circuit dataset. A QTask submission event is then scheduled according to its designated arrival time, and the QTask object is dispatched to the QBroker for processing. The QBroker subsequently acquires the current state of all QNodes and, integrating this with the incoming QTask's data, compiles an observation to relay to the RL actor. Utilizing Ray RLlib, the RL actor evaluates the observation and determines the appropriate action, which it communicates back to the QBroker. Following the action directive, the QBroker orchestrates task placement by forwarding the QTask to the specified QNode. Post-placement, the QBroker retrieves a new observation for the RL actor, which then assesses the reward and iteratively refines its policy, thereby enhancing its decision-making for future tasks.

\subsection{DRL Training Experiments and Results}
We implement the task scheduling policy using the Deep Q Network (DQN) \cite{mnih2015human} in QSimPy environment. We use Ray RLlib for implementing the policy with the hyperparameters set as follows: learning rate is 0.01; replay buffer capacity is 60000; the number of atoms is 10; the number of steps is 5; prioritized replay alpha, beta, and epsilon are set to 0.5, 0.5, and 3e-6, respectively. We evaluate the policy's performance over different numbers of training iterations based on episode reward values. We also observe the episode lengths of over 100 iterations, as these metrics reflect the number of task placements taken in each episode. 

\begin{figure}[htbp]
\centering
\includegraphics[width=2.6in]{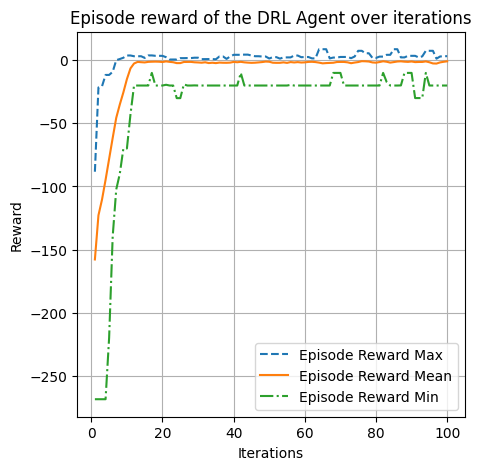}
\caption{Episode rewards of different iterations showed that the DRL-based task scheduling policy has been trained in QSimPy environment}
\label{fig_reward}
\end{figure}

\begin{figure}[htbp]
\centering
\includegraphics[width=2.6in]{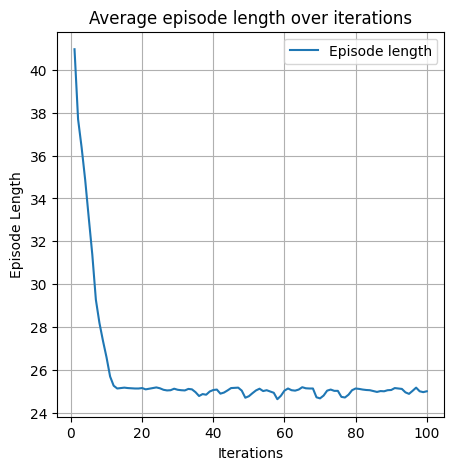}
\caption{Average of episode length has been reduced over time and reached the optimal value around 25 after 15 iterations}
\label{fig_len}
\end{figure}

Figure \ref{fig_reward} shows the episode reward values for the $10^5$ time steps taken over 100 iterations. Initially, the agent takes random actions, resulting in a lower negative reward as the incoming task cannot be executed successfully (for example, the scheduled QNode does not have enough qubits to execute QTask). After the agent learns and improves from previous experience,  it is clear that the average episode reward increases over time and converges after around 15 training iterations. Besides, the average number of episode lengths is reduced over time and reaches the optimal value of around 25. This result indicates that the number of violated actions (or rescheduling actions) decreases. Therefore, we have gained confidence that the agent has learned a policy that minimizes the total completion time, satisfying the initial quantum task scheduling objective.

\section{Conclusions and Future Work}
In this work, we proposed QSimPy, a novel discrete-event simulator and learning-centric environment architected for the rigorous demands of quantum cloud computing resource management problems. Its design harmoniously aligns with the needs of high-level, abstract modeling and practical, application-oriented research. By leveraging a Gymnasium-based environment and connecting with the Ray ecosystem for reinforcement learning integration, QSimPy allows for the creation and refinement of resource management policies that are adaptive, robust, and efficient. Our study confirms the framework's capability to support the design, development, and evaluation of learning-based quantum cloud resource management techniques.

Several promising directions for further research can be considered for extending QSimPy's capabilities. The framework's adaptability to evolving quantum hardware and algorithms lays the groundwork for extensive, real-world deployment scenarios. New enhancements can include support for distributed quantum computing models and incorporating quantum error correction simulations. Additionally, diversifying the range of reinforcement learning algorithms tested within the QSimPy environment can deepen our understanding of their efficacy in quantum task scheduling.

\vspace{12px}
\noindent \textbf{Software availability} \\ 
\noindent The QSimPy framework with the source code of proof-of-concept implementation are released on Github at (\href{https://github.com/Cloudslab/qsimpy }{github.com/Cloudslab/qsimpy}) as an open-source tool under the GPL-3.0 license.

\section*{Acknowledgment}
Hoa Nguyen acknowledges the support from the Science and Technology Scholarship Program for Overseas Study for Master’s and Doctoral Degrees, Vingroup, Vietnam.

\bibliographystyle{ieeetr}
\bibliography{bibliography}

\end{document}